\documentclass[
aps,%
11pt,%
final,%
titlepage,%
oneside,%
twocolumn,% 
nobibnotes,% 
nofootinbib,%
superscriptaddress,% 
showkeys,%
centertags]%
{revtex4}

\usepackage{amsmath}
\usepackage{graphicx}
\usepackage[colorlinks=true,linkcolor=blue,citecolor=blue,urlcolor=blue]{hyperref}

\usepackage{tabularx}

\usepackage{amssymb}

\setcitestyle{authoryear}

\usepackage{setspace}

\begin{document}

\singlespacing

\title{\LARGE The Hubble Diagram: Jump from Supernovae to~Gamma-Ray Bursts}

\author{\firstname{Nikita}~\surname{Lovyagin}}
\email{n.lovyagin@spbu.ru}
\affiliation{Saint Petersburg State University,
7/9 Universitetskaya Nab., St Petersburg 199034, Russia}

\author{\firstname{Rustam}~\surname{Gainutdinov}}
\email{roustique.g@gmail.com}
\affiliation{Saint Petersburg State University,
7/9 Universitetskaya Nab., St Petersburg 199034, Russia}

\author{\firstname{Stanislav}~\surname{Shirokov}}
\email{arhath.sis@yandex.ru}
\affiliation{SPb Branch of Special Astrophysical Observatory of Russian Academy of Sciences, 65 Pulkovskoye Shosse, St Petersburg 196140, Russia}

\author{\firstname{Vladimir}~\surname{Gorokhov}}
\email{vlgorohov@mail.ru}
\affiliation{Saint Petersburg Electrotechnical University, Ulitsa Professora Popova 5, 197376 St. Petersburg, Russia}

\begin{abstract} 
The Hubble diagram (HD) is a plot that contains luminous distance modulus presented with respect to the redshift. The distance modulus--redshift relation of the most well-known ``standard candles'', the type Ia supernovae (SN), is a crucial tool in cosmological model testing. In this work, we use the SN Ia data from the Pantheon catalogue to calibrate the Swift long gamma-ray bursts (LGRBs) as ``standard candles'' via the Amati relation. Thus, we expand the HD from supernovae to the area of the Swift LGRBs up to $z\sim8$. To improve the quality of estimation of the parameters and their errors, we implement the Monte-Carlo uncertainty propagation method. We also compare the results of estimation of the Amati parameters calibrated by the SN Ia, and by the standard $\Lambda$CDM model and find no statistically significant distinguish between them. Although the size of our LGRB sample is relatively small and the errors are high, we find this approach of expanding the cosmological distance scale perspective for future cosmological tests.
\end{abstract}
 % multimessenger observations--classical cosmological tests--cosmic tomography--

\newcommand{\orcidauthorA}{0000-0002-8386-7659} % Add \orcidC{} behind the author's name
\newcommand{\orcidauthorD}{0000-0002-8206-9890} % Add \orcidE{} behind the author's name

\maketitle

\newcommand{\di}{\mathrm{d}}
\newcommand{\keV}{\mathrm{keV}}
\newcommand{\erg}{\mathrm{erg}}
\newcommand{\scs}{\mathrm{s}}
\newcommand{\cm}{\mathrm{cm}}
\newcommand{\Ep}{E_{\mathrm{p}}}
\newcommand{\Epi}{E_{\mathrm{p,}i}}
\newcommand{\Emin}{E_{\mathrm{min}}}
\newcommand{\Emax}{E_{\mathrm{max}}}
\newcommand{\Eiso}{E_{\mathrm{iso}}}
\newcommand{\Sbolo}{S_{\mathrm{bolo}}}
\newcommand{\Sobs}{S_{\mathrm{obs}}}

\section{Introduction}

The Hubble diagram (HD) is a plot of object redshifts $z$ with respect to object distances $d$. 
The HD is a well-known and widely-used practical cosmological test~\citep{sandage1995astronimical,baryshev2012fundamental,shirokov2020high}.
Based on cosmological models, one usually describe theoretical the redshift--distance relation as a parametric function $d(z,\mathbf p)$, where $\mathbf p$ is a parameter vector. Thus, cosmology-independent determination of distances to objects with known redshift gives one a unique opportunity to verify, compare and probe  parameters of cosmological models. In case of the standard $\Lambda$CDM cosmological model, for the HD $\mathbf p$ may be considered, for example, as $(H_0, \Omega_M)$.

Usually, the HD is built up using the so-called standard candles (SC), i.e. the objects with theoretically or empirically known absolute brightness (or magnitude). Measured visible magnitudes of SCs directly give one required distances. At the end of the 20th century, the HD was constructed for Type Ia supernovae (SNe Ia) standard candles. Thus, the accelerated expansion of Universe within the standard Friedmann--Lemaitre--Robertson--Walker (FLRW)~\citep{baryshev2012fundamental} cosmological model was discovered. This led to the introduction of dark energy into the standard cosmological model (SCM) \citep{riess1998observational,perlmutter1999measurements}. However, there is still a wide discussion on cosmological models and their parameter values \citep{aghanim2020planck,riess2018new,riess2020expansion,yershov2020distant}. The $w$CDM model, where the $\Lambda$CDM model is a special case of one, is often considered as an alternative to SCM. The $w$CDM model is defined as the FLRW model that contains two cosmological non-interacting fluids, having equation of state of the cold matter $p = 0$, and the quintessence (dark energy) $p = w\rho c^2$ (with $w < 0$)~\citep{baryshev2012fundamental}.

The current limit of SN observations is about $z\sim{}$ 2--3, while the long gamma-ray bursts (LGRBs) up to $z\approx 10$ and are already seen~\citep{amati2018theseus, stratta2018theseus}. This makes them promising objects to prolong the HD significantly further into the Universe. The LGRB sources are related to explosions of massive core-collapse SN in distant galaxies~\citep{cano2017observer}, though up to now there is no satisfactory theory of the LGRB radiation origins~\citep{willingale2017gamma,fraija2021origin}. There are a number of studies that suggest using LGRBs as SCs~\citep{amati2002intrinsic,ghirlanda2004collimation,ghirlanda2007confirming,amati2008measuring,amati2019addressing,demianski2017cosmology,demianski2017cosmology2,lusso2019tension,yonetoku2004gamma,wang2011updated,wei2017gamma}. We suppose that LGRB HD can be used for probing cosmological parameters $\mathbf p$ and comparing cosmological models, as well as the SNe HD~\cite{shirokov2020high,Shirokov2020b,shirokov2020theseus}.

The Amati relation is an observed linear-like correlation between LGRB spectrum parameters, including redshift and distance to LGRB host galaxy, in logarithmic plot~\citep{amati2008measuring}. 
% In the case of known estimates of jet opening angle, a more physically correct the Ghirlanda relation can be used~\citep{}.
Despite the fact that individual LGRBs may not satisfy the Amati relation due to both physical and observational factors, we expect that for a statistically large ensemble of LGRBs, the correlation is on average correct. So the relation gives one an statistical opportunity to measure the distance to LGRBs with known redshift independently of a cosmological model. 
However, it depends on two unknown parameters that have to be calibrated observationally. For calibration of the correlation, the LGRBs with known distances and redshifts are needed so that the first step of our study is in determining distances $d$ for a subsample of LGRBs in near galaxies by cosmology-independent methods. 

In this study, we try to calibrate the Swift LGRBs~\footnote{\url{https://swift.gsfc.nasa.gov/archive/grb_table/}} as standard candles by using $\Lambda$CDM as a basis, i.e. obtaining distances from the model. However, this approach involves the circularity problem~\citep{kodama2008gamma} as for proper calibration of cosmologically-independent distance measurement should be used. So next we calibrate LGRBs as standard candles by using the Pantheon SNe Ia~\citep{scolnic2018complete} as a basis.

We decided to approximate the SN HD by using a smooth elementary mathematical function $d^{SN}(z)$ that can be directly used to get the distances to sources of LGRBs with known $z$. The LGRBs should be near enough, so they can lie inside representative SN sample, where the distance error $\sigma_{d^{SN}(z)}$ is small enough. Corresponding cosmological background of the Amati relation and introduction in mathematical approach of its calibration was described in~\citet{shirokov2020high}: in this paper, we apply the proposed idea of a cosmological model-independent calibration of gamma-ray bursts using a modified formula and advanced statistical methods.

The second step is in determining the Amati relation parameters via selected near LGRBs (i.e. LGRB calibrating), finding distances to all LGRBs with known redshift via the calibrated Amati relation, and plotting the HD for them. In fact, SNe are also calibrated SCs via cepheids that are also calibrated (via parallaxes) so that our approach lies in the frameworks of extending the cosmological distance scale. Since the supernova catalogues data are tied to a fixed value of the Hubble constant $H_0=70$ km/s/Mpc, this approach is not completely cosmologically-independent, all values are obtained with an accuracy of the scale factor $H_0$~\citep{hubble1929relation}.

At almost every stage of this study it is required to find best-fitting parameter estimation. In case, e.g., of the Amati parameters, the best-fitting function $y=f(x)$ should take into account errors in data points both of $x$ and $y$. Because of this and lack of LGRB statistics we find common linear least-square method to be unsuitable for our study. Instead we use the Theil-Sen estimator (also known as the single median method) for the linear regression purposes \citep{gilbert1987statistical}.
% Gilbert, Richard O. (1987), "6.5 Sen's Nonparametric Estimator of Slope", Statistical Methods for Environmental Pollution Monitoring, John Wiley and Sons, pp. 217–219, ISBN 978-0-471-28878-7
In this method, the estimation for the slope is defined as the median slope among all of the possible pairs of dots. We also need a method to perform the parameter estimation (the curve fitting) routines for the arbitrary function $f$. For this we use the trust region reflective algorithm~\citep{trustreg}, which is the non-linear least-squares method implemented in the \texttt{curve\_fit} function of the \texttt{scipy.optimize} Python library~\citep{SciPy}. 
We can interpret both of these methods as pipelines that return the estimated parameters based on the input data. Our purpose is to also estimate the errors and covariances of the parameters. To do this, we utilise the Monte-Carlo sampling uncertainty propagation approach~\citep{montecarloold, montecarlonew}. 
Applied to our pipelines, this approach allows us to correctly propagate the errors directly from the input data to the estimated parameters. 

In the paper we have calibrated the LGRBs from the selected subsample as SCs via supernovae, as it is described above, and have built the corresponding LGRB HD, which is the main result of this work. Nowadays, there are still not enough statistics and theory to get reliable values of cosmological parameters from the LGRB HD. However, we find our method perspective and promising one. We also compare $\Lambda$CDM-based and SN-based Amati parameters and find no significant difference between them.

\section{Materials and Methods}

\subsection{The Hubble diagram as a basic cosmological test}

The HD is the redshift--distance relation, which can be represented as a table of points $(z_i,d_i)$ on the plot for each $i$-th object. Usually, the luminosity distance $d_L$ is used. This value depends on the visible magnitude $m$, absolute magnitude $M$, and distance modulus $\mu$ as 
\begin{equation}
    \label{mu}m-M=\mu=5\log (d_L / 1\,\mathrm{Mpc})+25 \, .
\end{equation} 
In fact, a HD is plotting the $(z_i,\mu_i)$ relation. 
% plot of

In general, cosmological models give theoretical assumptions on the function $d_L(z, \mathbf p)$, where $\mathbf p$ are model parameters. For example, in $w$CMD model $\mathbf p = \{H_0,\,w,\,\Omega_w,\,\Omega_k\}$, and in $\Lambda$CMD model $\mathbf p = \{H_0,\,w=-1,\,\Omega_w=\Omega_\Lambda,\,\Omega_k=0.0\}$. The HD is one of the cosmological tests that allow one to find the model parameters via fitting the observational data. For this one, the redshift $z$ and distance $d_L$ should be determined independently (e.g., $z$ directly via spectrum, and $d_L$ via various indirect methods). 

\subsection{Gamma-ray bursts as standard candles}

Standard candles are a group of objects with the known (theoretically or empirically) typical absolute magnitude $M$. They allow one to find luminosity distance or distance modulus directly from Equation~(\ref{mu}) by using the visible magnitude $m$, or bolometric fluxes.
		
In case of LGRBs, the situation is a little different. The Amati relation for LGRBs~\citep{amati2002intrinsic,amati2008measuring,amati2019addressing} is the equation 
\begin{equation}
    \log E_\mathrm{iso} = a\log E_{\mathrm{p},i}+b \, , 
\end{equation}
where 

\noindent $E_{\mathrm{p},i}=E_\mathrm{p}(1+z)$ is the rest frame spectral peak energy, where $E_\mathrm{p}$ is the observed LGRB parameter determined via fitting the LGRB spectrum by the cut-off power law (CPL) model function $N(E)$,
	\begin{equation*}
	N(E)\left[\frac{\mathrm{photons}}{\mathrm{KeV}\,\mathrm{s}\,\mathrm{cm}^2}\right] = AE^\alpha \exp \big( -(2 + \alpha)E/ E_\mathrm{p} \big) \, ,
	\end{equation*}
	where $\alpha$ and $E_\mathrm{p}$ are the CPL model parameters%~\citep{band1993batse} 
	of a LGRB from the Swift database;

\begin{figure}		
    \centering \includegraphics[width=\linewidth]{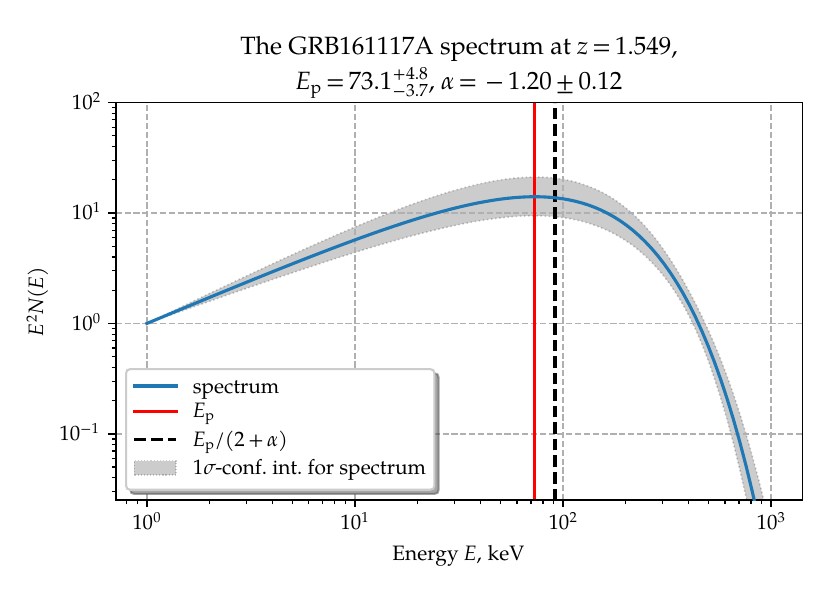}
    \caption{LGRB\,161117A spectrum in CPL model as an example.}
    \label{fig1:band}
\end{figure}

\noindent $E_\mathrm{iso}=4\pi d_L^2 \cdot S_\mathrm{bolo} / (1+z)$ is the isotropic equivalent radiated energy in gamma-rays~\citep{ajello2019decade}. The distance $d_L$ and observed integral fluence $S_\mathrm{bolo}$ are determined as quantities transferred per a unit energy frame area and that are corrected for the instrumental (observed) spectral energy range, and source redshift. The correction is performed by the equation
	\begin{equation*}
	S_\mathrm{bolo}= S_\mathrm{obs} \dfrac{\int_\frac{1}{1+z}^\frac{10^4}{1+z} EN(E)\;\mathrm{d}E }{\int_{E_\mathrm{min}}^{E_\mathrm{max}} E N(E)\;\mathrm{d}E} \, , 
	\end{equation*}
	where $S_\mathrm{obs}$ is the observed fluence, and $\{E_\mathrm{min},\,E_\mathrm{max}\}$ is the instrumental spectral energy range, which is $\{15, 150\}\,\mathrm{keV}$ for the Swift observatory; normalised example of such spectra is shown in Fig.~\ref{fig1:band};

\noindent $a$ and $b$ are the Amati relation parameters mentioned above that can be calibrated empirically as in this study.

If observed parameters $z$, $S_\mathrm{bolo}$, $E_{\mathrm{p},i}$, and model-independent measurement of $d_L(z)$ (at the small scales $z\lesssim 1.5$) are known, then best fit of the Amati relation parameters $a$ and $b$ can be found. 
Using $a$, $b$, $z$, $S_\mathrm{bolo}$, and $E_{\mathrm{p},i}$, one can find the LGRB distances $d^\mathrm{LGRB}_L(z)$ via the Amati relation and plot the LGRB HD. 

So, to find $a$ and $b$, one needs to get luminosity distances $d_L$ for near LGRBs. Although taking $d_L(z)$ from $\Lambda$CMD model or from any other cosmological model is a dead-loop (a circularity problem), we have first calibrated $a$ and $b$ using $\Lambda$CMD model using parameters from~\cite{aghanim2020planck} as a comparison basis.

Then we have chosen the way of the SN calibration. Thus, we attempted to build a function $d_L^\mathrm{SN}(z)$ by using SNe Ia as SCs. Distance $d_L$ was calculated as function $d_L^\mathrm{SN}(z)$ for near LGRBs that have redshifts $z$ low enough in order to perform an interpolation of the SN HD. 

This lets one to directly get the Amati relation parameters and to use them for cosmological model-independent determination of distances $d_L$ to all LGRBs with known observational data for $z$, $S_\mathrm{bolo}$, and $E_{\mathrm{p},i}$ and to plot the HD for them.

\subsection{Catalogues of SNe Ia and LGRBs}

We have used the Pantheon database (containing 1,048 SNe). In this catalogue, there are two kinds of redshift, which are the redshift in terms of the cosmic microwave background radiation $z_\mathrm{cmb}$, and the redshift in terms of the heliocentric system $z_\mathrm{hel}$ \citep{scolnic2018complete}. Between these two methods, we have chosen the $z_\mathrm{cmb}$ one. All the catalogue data is calculated in assumption of $H_0=70$ km/s/Mpc, so we have set this value for all our study.

For our purposes, we use the sample of 174 GRBs from the Swift catalogue with the measured parameters of $(\alpha,\,\Ep,\,\Sobs,\,z)$, where the spectral parameter $\alpha$, the peak energy $\Ep$, and the observed flux $\Sobs$ are described above, and $z$ is the redshift. Some of the uncertainties of these parameters were missing (several $\alpha$ errors and approximately a third of $\Ep$ errors were not presented). To deal with this issue, we assume the points with unknown errors to have the relative uncertainties equal to the median relative uncertainties of the corresponding parameter of the sample. The redshift $z$ values are also presented in the lack of uncertainties, as is expected. A significant proportion of the LGRB parameters has asymmetric and relatively huge errors, which leads to the inexpediency of using the common uncertainty propagation rule. Instead, we use the approach of Monte-Carlo sampling, which is thoroughly described in the Section~\ref{App:uncertainties}. Our LGRB sample is represented in Google Table\footnote{\url{https://docs.google.com/spreadsheets/d/1jbOQxUlwEYd8qWU68mbih_PBPOHp0iUs-3kgphUIvAs/edit?usp=sharing}}.

\subsection{Monte-Carlo uncertainty propagation}
\label{App:uncertainties}

The standard approach to the error propagation problem is known as the linear uncertainty propagation (LUP) theory. One of the best implemented error propagation software is the uncertainties package of the Python language. The errors in this approach are interpreted as the standard deviations, while the values have the sense of the mean. 
So, the values of variable with its error defines by the normal distribution, which denotes the possible whereabouts of the variable via distribution parameters. 
However, the LUP theory requires that the errors be relatively small compared to the values of variable, so that the functions in calculations are nearly linear compared to these small shifts. 
Also, the LUP approach is not capable of handling asymmetric errors. Thus, we cannot use it.\par

For that reason, we decided to use the Monte-Carlo sampling. In this approach, the values and their errors also have the sense of defining the distributions of possible locations, and we use these distributions to draw samples of the size of 10,000. To handle the asymmetric uncertainties, we interpret the values as the medians, the lower bounds are interpreted as the 0.16 quantiles, and the upper bounds as the 0.84 quantiles. With this interpretation of errors, the case of symmetric uncertainties reduces to the normal distribution with the known mean and standard deviation. In the case of asymmetric errors, we use the split-normal distribution, which results from joining the two halves of normal distributions with different standard deviations at their mode. Some of the $\Ep$ values from our LGRB data set had remarkably huge lower uncertainties, so that trying to Monte-Carlo sample them would lead to negative $\Ep$ values, which has absolutely no physical sense. Because of that, in the case of $\Ep$ variables, we draw the split-normal distributions in the space of $\log \Ep$. So, the peak energy values are drawn using the log-split-normal distributions. Taking the logarithms does not move the quantiles, so the errors in this case are not changed. 

The Monte-Carlo approach to propagating the errors is simple yet powerful. It allows one not only to calculate the uncertainties of calculated values, but also to track and take into account the correlations between variables for free.

\subsection{Best-fitting methods}

To obtain the best-fit parameters of the approximation function $d_L^{SN}(z)$ for the SN HD, one needs to minimise the functional value
\begin{equation}
    \label{chi}
    \chi^2=\sum_{i=1}^n \frac 1{\sigma_i^2} \bigl(y_i-f(x_i,\mathbf p)^2\bigr) \, ,
\end{equation}
where 
\begin{itemize}
	\item $(x_i,y_i)$ are observed table values;
	\item $\sigma_i$ is error of $i$-th value;
	\item $f$ is the model function and $\mathbf p = \{p_1,p_2,\dots,p_m\}$ are the parameters.
\end{itemize}

% In this work we use the \texttt{curve\_fit} function implemented in the \texttt{scipy.optimize} Python library~\citep{SciPy}. 
This function utilizes the trust region reflective algorithm~\citep{trustreg} to minimize the $\chi^2$ function.

\subsection{Interpolation function of the SN HD}

In fact, we can use any smooth function to use it as a $d_L^{SN}(z)$ function. 
All that we need in order to minimize the error between the real value and a mathematically predicted one for the luminous distance at any low $z$, where we have enough supernovae to do this, is the approximation accuracy that is provided by our approach in the redshifts $z_{min} \ll z\lesssim z_{max}$, where $z_{min}$ and $z_{max}$ are redshifts of the nearest and the farthest supernovae ($z=0.01012$ and $z\approx 2.26$, respectively). 
% Of course, any extrapolation using this function would have given an incorrect (non-precious) result.

\begin{figure}[h]	
	\centering \includegraphics[width=\linewidth]{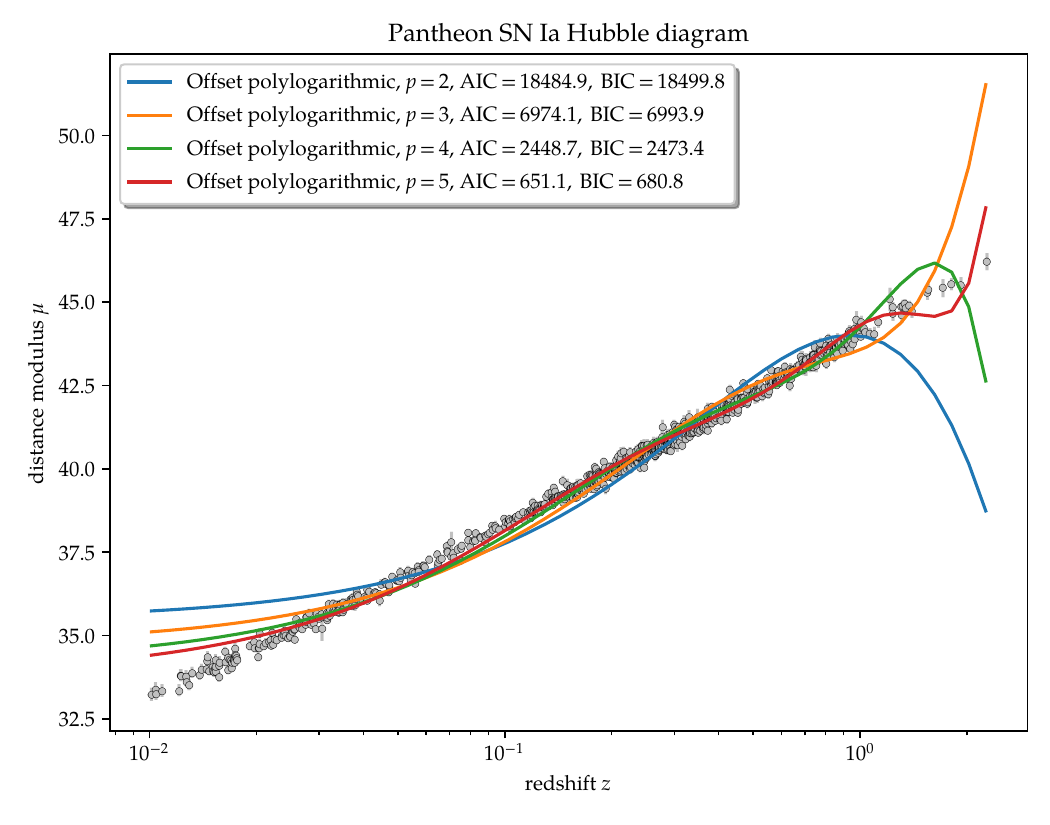}
	\includegraphics[width=\linewidth]{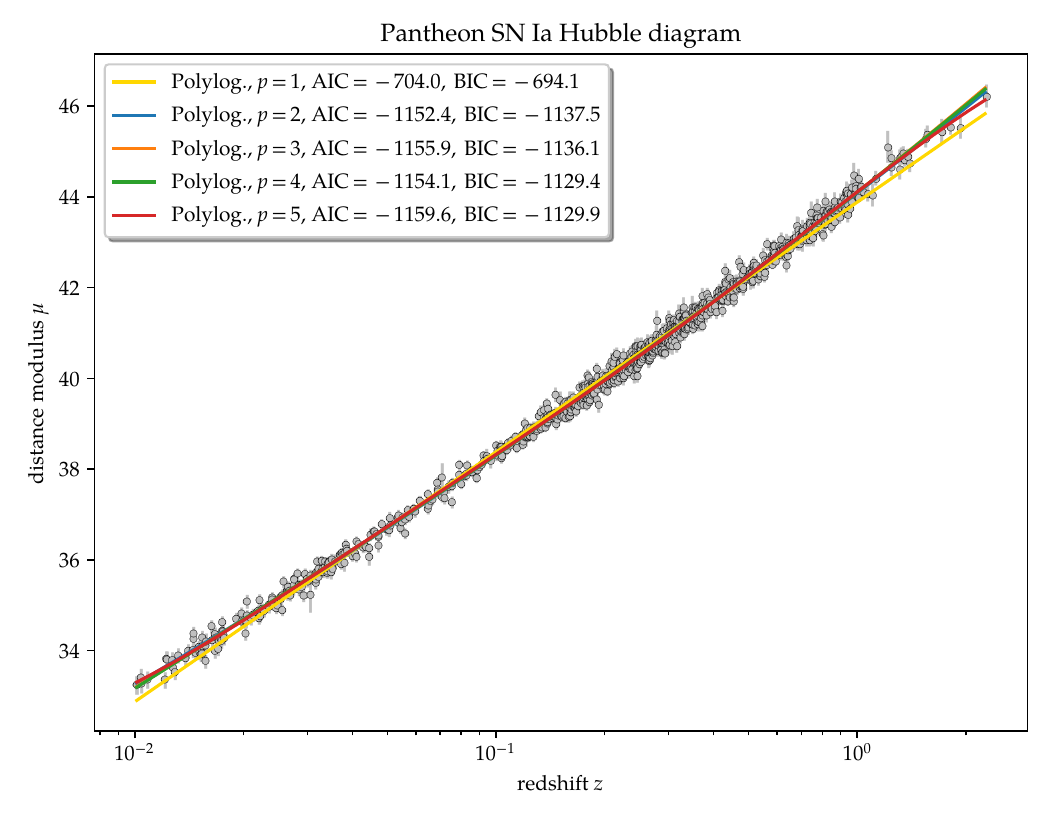}
	
	\caption{The result of fitting the shifted polylogarithmic function (top) and polylogarithmic function (bottom) to the Pantheon SNe Ia data.}
	\label{figSNbad:snp}
\end{figure}

Since in logarithmic scales (along both of the distance and redshift) the relation $d_L(z)$ is already known, it has almost linear behaviour at low $z$, we have decided to use a polylogarithmic function of a degree $k$. We have tried to use three following functions:
\begin{itemize}
    \item theoretically-inspired function 
    \begin{equation}\label{tj}\mu^{SN}(z) = 5\log\frac{cz}{H_0}+25+\sum\limits_{i=1}^p a_i\log^i(1+z);
    	\end{equation}
    \item simple polylogarithmic function     \begin{equation}\label{dlsn}\mu^{SN}(z) = \sum\limits_{i=0}^p a_i\log^iz;
    	\end{equation}
    \item shifted polylogarithmic function     \begin{equation}\mu^{SN}(z) = \sum\limits_{i=0}^p a_i\log^i(1+z).
    	\end{equation}
\end{itemize}
The first function may be considered as addition of small correction to the linear Hubble law at low $z$. The results of using other function is shown in Fig. \ref{figSNbad:snp}.

\section{Results}

\subsection{Approximation of the SN HD}

\begin{figure*}
    \centering \includegraphics[width=\linewidth]{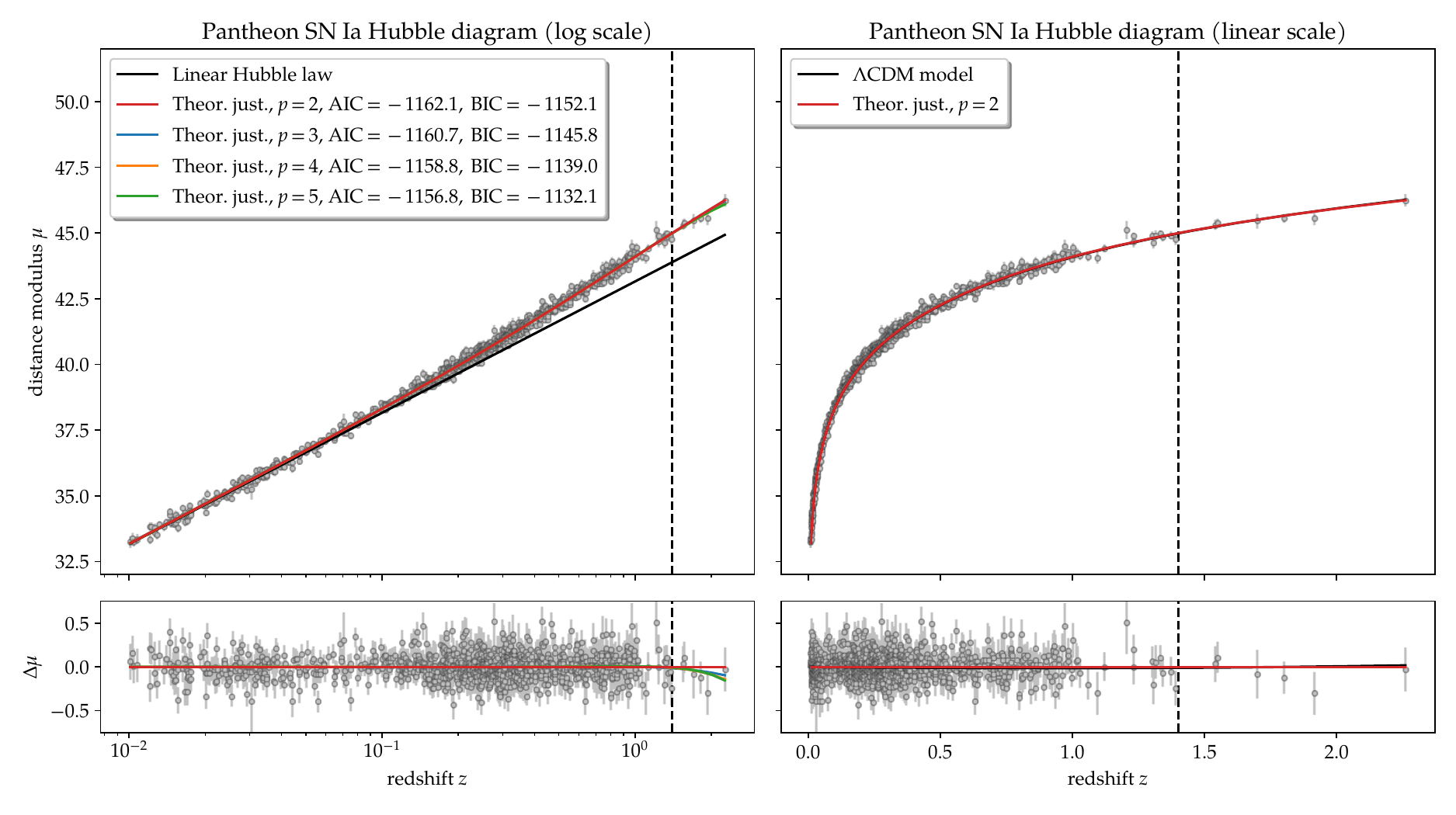}
    \caption{The result of fitting the Pantheon SNe Ia data by the theoretically justified function. The model with $p=2$ is chosen as the best one due to minimal Akaike Information Criterion (AIC) value and minimal Bayesian Information Criterion (BIC) value. The vertical dashed line marks the redshift of $z=1.4$, which is the upper bound for the near LGRBs used to calibration. The interpolated distance--redshift relation would be used only for the LGRBs up to this border.}
    \label{fig2:snp}
\end{figure*}

The polynomials given by equation~(\ref{dlsn}) for degrees $p=1,\,2,\,3,\,4,\,5$ have been used. The result of best-fitting is shown in figure~\ref{fig2:snp}. 

\begin{figure}		
    \centering \includegraphics[scale=0.75]{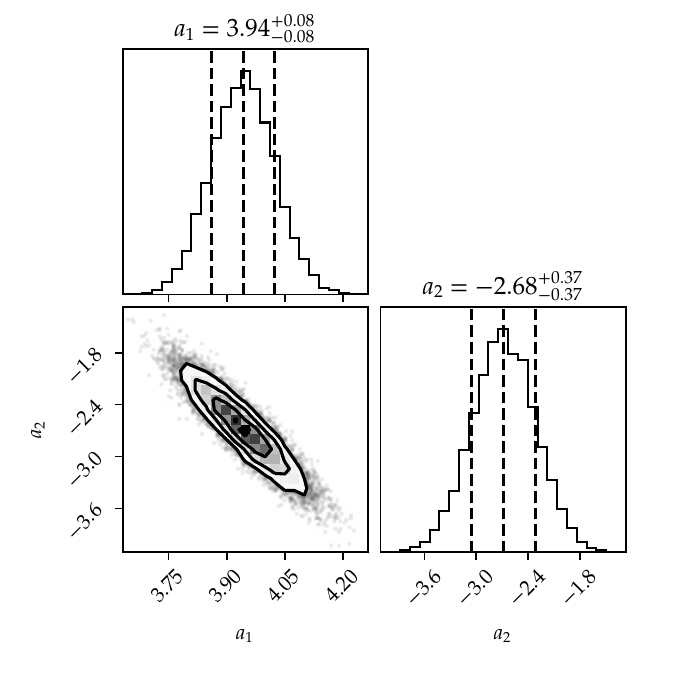}
    \caption{The sample of best-fit parameters for the theoretically justified model given by the Eq.~(\ref{tj}) with $p=2$ drawn from the multivariate normal distribution with the size of 10,000 via the obtained estimates of parameters and their covariance matrix.}
    \label{fig3:corner_pl3}
\end{figure}

Usually, polynomials of degree 1 or degree 2 are used for HD approximation~\citep[e.g.,][]{amati2019addressing}. We have chosen the model with $p=2$, because it minimises the Akaike Information Criterion (AIC)~\citep{akaike1974new}. We use the obtained estimates of the parameters and their covariance matrix to generate them from a multivariate normal distribution of the size of 10,000. The corner plot of this sample is shown in the figure~\ref{fig3:corner_pl3}. 
This sample will come useful later for the Monte-Carlo uncertainty propagation. 
These polynomial coefficients have no physical sense, they just show the best-fit of a cosmology received by model-independent redshift--distance relation.
 
\subsection{Amati relation parameters probing and Gamma-Ray Bursts Hubble Diagram}

By analysing figure~\ref{fig2:snp}, we took LGRBs with $z<1.4$ to calibrate the Amati coefficients $a$, and $b$. All SNe samples in this range are representative enough and their approximations via polylogarythmic function have low values of its formal errors. In total, 75 of 174 LGRBs in this range (with known $z$ and calculable $E_{\text{p},i}$ and $S_\text{bolo}$) are available. To estimate the Amati parameters $a$ and $b$, we use the Theil-Sen estimation, which is a reliable and robust method for linear regression. In this method, the slope (parameter $a$) is estimated as the median of all of the slopes between all pairs of points. The intersect (parameter $b$) is then estimated as the median of the values $y_i - ax_i$. Thus, we have the pipeline that takes the LGRBs data table and returns the Amati coefficients $a$ and $b$. The Monte-Carlo error propagation approach can be also applied to this pipeline, so that the output would consist not only of the estimated parameters of $a$ and $b$, but their Monte-Carlo samples of the size of 10,000. We can further use the medians of this samples as the estimated values for the parameters, and the quantiles of 0.16 and 0.84 as the upper and lower 1$\sigma$-borders. 

\begin{figure*}
    \centering 
    \includegraphics[width=0.54\linewidth]{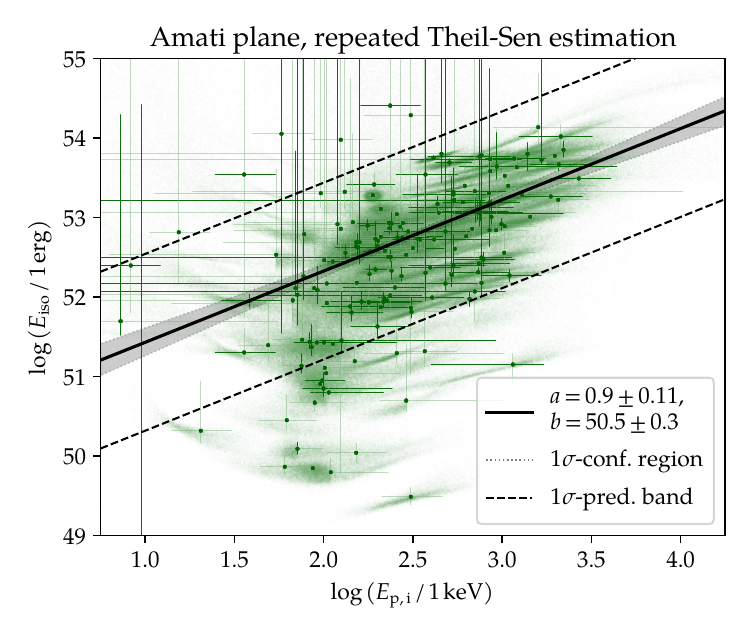} 
    \includegraphics[width=0.44\linewidth]{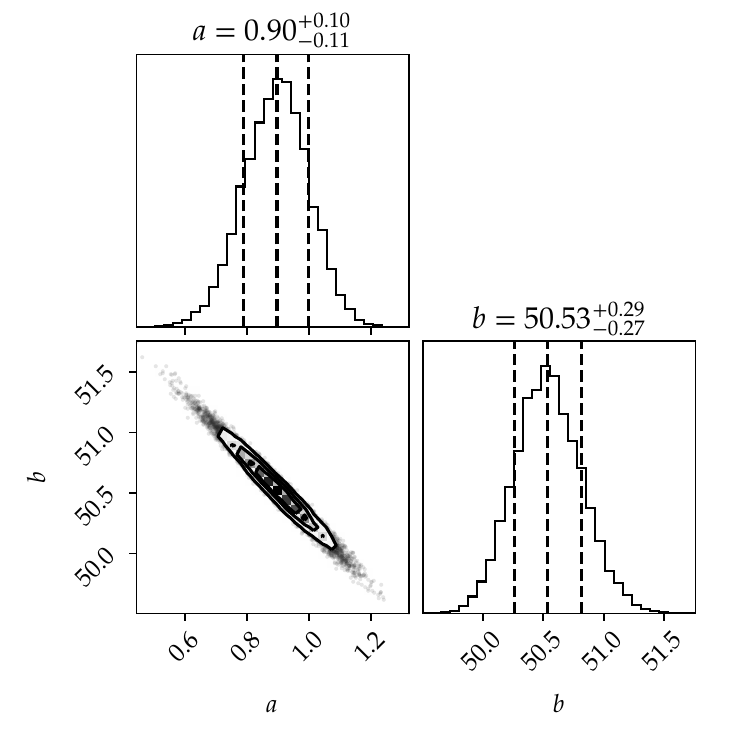} \\
    \includegraphics[width=0.54\linewidth]{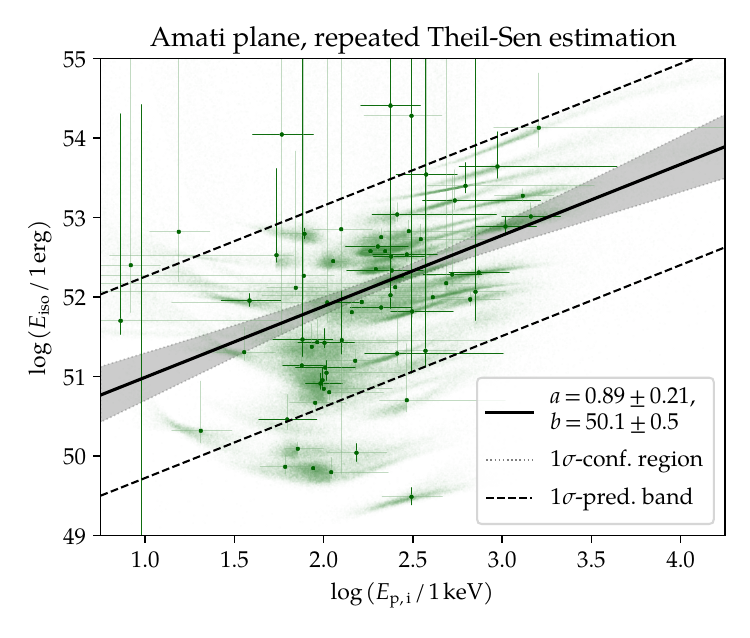} 
    \includegraphics[width=0.44\linewidth]{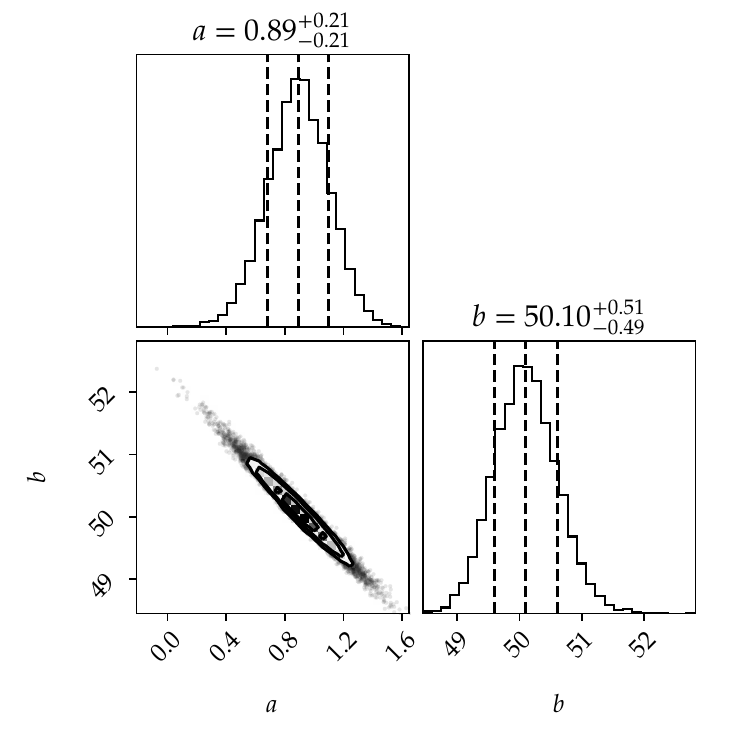}
\caption{\textit{Left}: the plane of the $\log E_{\mathrm{iso}}$ -- $\log E_{\mathrm{p},i}$ parameters (the Amati plane). Each LGRB is represented by the point with the error bars and the underlying ``Monte-Carlo cloud'' of the size of 10,000. The plot also shows the line and the confidence region that corresponds to the estimated $a$ and $b$ parameters with their errors and covariance.
\textit{Right}: the Monte-Carlo sample corner plot for the Amati parameters $a$ and $b$. The error borders are defined as the 0.16 and 0.84 quantiles.
\textit{Top}: all 174 LGRBs, distances taken from $\Lambda$CDM.
\textit{Bottom}: subsample of 75 LGRBs with $z < 1.4$, distances taken from SN HD theoretically justified approximation function with $p=2$.
}
\label{fig4:AmatiSN}
\end{figure*}

\begin{figure*}
    \centering
    \includegraphics[scale=0.75]{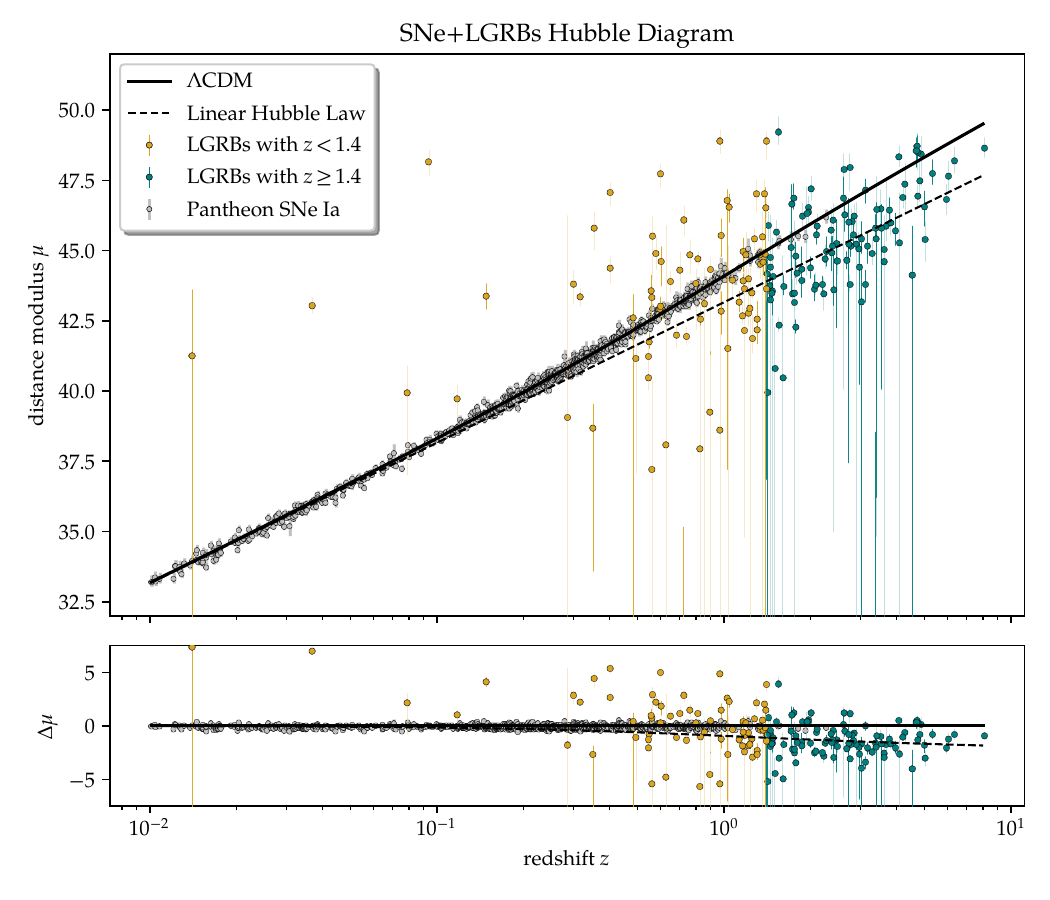}
    \caption{The Hubble diagram for our sample of 174 LGRBs together with the type Ia SNe form the Pantheon catalogue.}
    \label{fig5:hd}
\end{figure*}

The results of the estimation process are presented in the figure~\ref{fig4:AmatiSN}. "Repeated Theil-Sen estimation" means that the figure shows the average regression of 10,000 ones. In Table~\ref{tab:ab}, we compare the Amati parameters, calculated through $d_L^{\Lambda CDM}$ and $d_L^{SN}$, where $d_L^{\Lambda CDM}$ calculated by equation (B5) from~\citet{shirokov2020high}. With the estimated Amati parameters $a$ and $b$, it is now possible to find the distance modulus for the whole sample of LGRBs. The final SN+LGRB HD is shown in Fig.~\ref{fig5:hd}.

\begin{table}[]
    \caption{Calibration results of the Amati parameters for all LGRBs calibrated by the $\Lambda$CDM model and for the near LGRBs calibrated by SN Ia.}
    \label{tab:ab}
    \centering
    \begin{tabular}{c|c|c}
    \hline
        Amati parameter & $a$ & $b$ \\ \hline
        Value from $\Lambda$CDM & $0.90\pm0.11 $ & $50.5\pm0.3 $ \\
        Value from SN & $0.89\pm 0.21 $ & $50.1\pm 0.5 $ \\ \hline \hline
    \end{tabular}
\end{table}

%%%%%%%%%%%%%%%%%%%%%%%%%%%%%%%%%%%%%%%%%%
\section{Discussion and conclusion}

We have calibrated the near Swift LGRB sample up to $z<1.4$ as standard candles through the Pantheon SN Ia catalogue using the Amati relation. The calibrated HD up to $z\sim 7-8$ is shown in figure~\ref{fig5:hd}. This diagram is constructed naturally from the standard cosmological model at fixed value of $H_0=70$~km/s/Mpc, and it does not take into account any systematic corrections.

The Amati relation for the near LGRBs calibrated by SN Ia and for all LGRBs calibrated by the $\Lambda$CDM model (Table~\ref{tab:ab}) match with $1\sigma$ level. It can be concluded that, in terms of statistic significance, there are no observed deviations from the standard cosmological model for the far LGRBs with $z>1.4$.

However, visual analysis of the HD in the figure~\ref{fig5:hd} shows a trend towards the fogging of distant LGRBs relatively to the standard model (shifting to the linear Hubble law). Possible explanations include LGRB evolution~\citep{wiseman2017evolution,butler2007x}; gravitational lensing~\citep{veres2021fermi} and gravitational mesolensing that leads into increasing the brightness of objects in all spectral ranges\citep{baryshev2002gravitational,raikov2016ultraluminous,raikov2021superluminous}, observational selection, and others. It should also be noted that the cosmic time dilation effect is taken into account in this study. This implies that our results are valid within models that include this effect. These issues require additional research, including the accumulation of the LGRB sample and the development of statistical methods.

The GRB observations in multimessenger astronomy epoch open new possibilities for testing the fundamental physics lying in the basis of the standard cosmological model: classical general relativity, cosmological principle of matter homogeneity, and the
Lemaıtre space expansion nature of cosmological redshift.

The corrected HD cosmological tests can probe strong-field regime of gravitation theory, spatial distribution of galaxies, Hubble Law and time dilation of physical processes at high redshifts. Constructing the high-redshift GRB Hubble diagram and comparison of time dilation in GRB pulses,
GRB afterglow and core-collapse SN light curves to
test the expanding space paradigm~\citep{fraija2020grb}.
Perspectives for performing cosmological tests in multimessenger astronomical observations of GRBs were considered and several new tests were proposed in~\citet{shirokov2020theseus, Shirokov2020b}. 
The LGRB HD is the test that could be used, in particular, to gravitational lensing and Malmquist biases testing~\citep{shirokov2020high}, classical general relativity, cosmological principle of matter homogeneity, and the Lemaitre space expansion nature of cosmological redshift testing. The GRB HD can be combined with the gravitational wave standard sirens at intermediate redshifts~\citep{schutz1986determining,holz2005using,abbott2017gravitational}. 
We find the method of LGRBs HD construction proposed in this paper perspective for performing such tests in future.

\vspace{6pt} 

\subsection*{Author contributions}
Conceptualization, Nikita Lovyagin and Rustam Gainutdinov; Formal analysis, Rustam Gainutdinov; Funding acquisition, Stanislav Shirokov; Investigation, Stanislav Shirokov; Methodology, Nikita Lovyagin, Rustam Gainutdinov, Stanislav Shirokov and Vladimir Gorokhov; Project administration, Stanislav Shirokov; Resources, Rustam Gainutdinov; Software, Rustam Gainutdinov; Supervision, Stanislav Shirokov and Vladimir Gorokhov; Validation, Nikita Lovyagin and Stanislav Shirokov; Visualization, Rustam Gainutdinov; Writing -- original draft, Nikita Lovyagin; Writing -- review \& editing, Vladimir Gorokhov.

\subsection*{Funding}
Part of the observational data was exposured on the unique scientific facility the Big Telescope Alt-azimuthal SAO RAS and the data processing was supported under the Ministry of Science and Higher Education of the Russian Federation grant 075-15-2022-262 (13.MNPMU.21.0003)

\subsection*{Data availability}
The codes developed in Python underlying this article are available in the repository on \url{https://github.com/Roustique/sngrb} 

\subsection*{Acknowledgments}
We thank the anonymous reviewer for important suggestions that helped us to improve the presentation of our results.

\subsection*{Conflict of interest}
The authors declare no conflict of interest.

\subsection*{Abbreviations}
The following abbreviations are used in this manuscript:\\

\noindent 
\begin{tabular}{@{}ll}
LGRB(s) & Long Gamma-Ray Burst(s)\\
HD & Hubble diagram\\
$\Lambda$CDM & $\Lambda$ Cold Dark Matter\\
SN(s) & Supernova(e)\\
SC(s) & Standard Candle(s) \\
\end{tabular}

\clearpage

\bibliographystyle{my_arXiv}
\bibliography{article.bib}

\end{document}